# Nonlinear magnetization dynamics of antiferromagnetic spin resonance induced by intense terahertz magnetic field

Y Mukai[1,2,4,6], H Hirori[2,3,4,7], T Yamamoto[5], H Kageyama[2,5], and K Tanaka[1,2,4,8] 

[1] Department of Physics, Graduate School of Science, Kyoto University, Sakyo-ku, Kyoto 606-8502, Japan

[2] Institute for Integrated Cell-Material Sciences (WPI-iCeMS), Kyoto University, Sakyo-ku, Kyoto 606-8501, Japan

[3] PRESTO, Japan Science and Technology Agency, Kawaguchi, Saitama 332-0012, Japan

[4] CREST, Japan Science and Technology Agency, Kawaguchi, Saitama 332-0012, Japan

[5] Department of Energy and Hydrocarbon Chemistry, Graduate School of Engineering, Kyoto University, Nishikyo-ku, Kyoto 615-8510, Japan

**E-mail:**

[6] mukai@scphys.kyoto-u.ac.jp

[7] hirori@icems.kyoto-u.ac.jp

[8] kochan@scphys.kyoto-u.ac.jp

We report on the nonlinear magnetization dynamics of a $HoFeO_3$ crystal induced by a strong terahertz magnetic field resonantly enhanced with a split ring resonator and measured with magneto-optical Kerr effect microscopy. The terahertz magnetic field induces a large change (~40%) in the spontaneous magnetization. The frequency of the antiferromagnetic resonance decreases in proportion to the square of the magnetization change. A modified Landau-Lifshitz-Gilbert equation with a phenomenological nonlinear damping term quantitatively reproduced the nonlinear dynamics.

## 1. Introduction

Ultrafast control of magnetization dynamics by a femtosecond optical laser pulse has attracted considerable attention from the perspective of fundamental physics and technological applications of magnetic recording and information processing [1]. The first observation of subpicosecond demagnetization of a ferromagnetic nickel film demonstrated that a femtosecond laser pulse is a powerful stimulus of ultrafast magnetization dynamics [2], and it has led to numerous theoretical and experimental investigations on metallic and semiconducting magnets [3-8]. The electronic state created by the laser pulse has a strongly nonequilibrium distribution of free electrons, which consequently leads to demagnetization or even magnetic reversal [1,2,9-11]. However, the speed of the magnetization change is limited by the slow thermal relaxation and diffusion, and an alternative technique without the limits of such a thermal control and without excessive thermal energy would be desirable.

In dielectric magnetic media, carrier heating hardly occurs, since no free electrons are present [12]. Consequently, great effort has been devoted to clarifying the spin dynamics in magnetic dielectrics by means of femtosecond laser pulses. A typical method for nonthermal optical control of magnetism is the inverse Faraday effect, where circularly polarized intense laser irradiation induces an effective magnetic field in the medium. Recently, new optical excitation methods avoiding the thermal effect such as the magneto-acoustic effect is also reported [13,14]. In particular, these techniques have been used in many studies on antiferromagnetic dielectrics because compared with ferromagnets, antiferromagnets have inherently higher spin precessional frequencies that extend into the terahertz (THz) regime [12,15]. Additionally, ultrafast manipulation of the antiferromagnetic order parameter may be exploited in order to control the magnetization of an adjacent ferromagnet through the exchange interaction [16]. The THz wave generation technique is possibly a new way of optical spin control through direct magnetic excitation without undesirable thermal effects [17-19]. As yet however, no technique has been successful in driving magnetic motion excited directly by a magnetic field into a nonlinear dynamics regime that would presumably be followed by a magnetization reversal [20-22].

In our previous work [23], we demonstrated that the THz magnetic field can be resonantly enhanced with a split ring resonator (SRR) and may become a tool for the efficient excitation of a magnetic resonance mode of antiferromagnetic dielectric $HoFeO_3$. We applied a Faraday rotation technique to detect the magnetization change but the observed Faraday signal averaged the information about inhomogeneous magnetization induced by localized THz magnetic field of the SRR over the sample thickness [23]. In this Letter, we have developed a time-resolved magneto-optical Kerr effect (MOKE) microscopy in order to access the extremely field-enhanced region, sample surface near the SRR structure. As a result, the magnetic response deviates from the linear response in the strong THz magnetic field regime, remarkably showing a redshift of the antiferromagnetic resonance frequency that is proportional to the square of the magnetization change. The observed nonlinear dynamics could be reproduced with a modified Landau-Lifshitz-Gilbert (LLG) equation having an additional phenomenological nonlinear damping term.

## 2. Experimental setup

Figure 1 shows the experimental setup of MOKE microscopy with a THz pump pulse excitation. Intense single-cycle THz pulses were generated by optical rectification of near-infrared (NIR) pulses in a $LiNbO_3$ crystal [24-26]; the maximum peak electric field was 610 kV/cm at focus. The sample was a $HoFeO_3$ single crystal polished to a thickness of 145 μm, with a c-cut surface in the *Pbnm* setting [27]. (The x-, y-, and z-axes are parallel to the crystallographic a-, b-, and c-axes, respectively.) Before the THz pump excitation, we applied a DC magnetic field to the sample to saturate its magnetization along the crystallographic c-axis. We fabricated an array of SRRs on the crystal surface by using gold with a thickness of 250 nm. The incident THz electric field, parallel to the metallic arm with the SRR gap (the x-axis), drove a circulating current that resulted in a strong magnetic near-field normal to the crystal surface [23,28,29]. The SRR is essentially subwavelength LC circuit, and the current induces magnetic field $\boldsymbol{B}_{nr}$ oscillating with the LC resonance frequency (the Q-factor is around 4). The right side of the inset in figure 1 shows the spatial distribution of the magnetic field of the SRR at the LC resonance frequency as calculated by the finite-difference time-domain (FDTD) method. Around the corner the current density in the metal is very high, inducing the extremely enhanced magnetic field in the $HoFeO_3$ [29].

At room temperature, the two magnetizations $\boldsymbol{m}_i$ (i=1,2) of the different iron sublattices in $HoFeO_3$ are almost antiferromagnetically aligned along the x-axis with a slight canting angle $\beta_0$(=0.63°) owing to the Dzyaloshinskii field and form a spontaneous magnetization $\boldsymbol{M}_S$ along the z-axis [30]. In the THz region, there are two antiferromagnetic resonance modes (quasiantiferromagnetic (AF) and quasiferromagnetic (F) mode [31]). The magnetic field $\boldsymbol{B}_{nr}$ generated along the z-axis in our setup causes AF-mode motion; as illustrated in figure 2(a), the Zeeman torque pulls the spins along the y-axis, thereby triggering precessional motions of $\boldsymbol{m}_i$ about the equilibrium directions. The precessional motions cause the macroscopic magnetization $\boldsymbol{M}=\boldsymbol{m}_1+\boldsymbol{m}_2$ to oscillate in the z-direction [32,33]. The resultant magnetization change $\Delta M_z(t)$ modulates the anti-symmetric off-diagonal element of the dielectric tensor $\varepsilon^{a}_{xy}(= -\varepsilon^{a}_{yx})$ and induces a MOKE signal (Kerr ellipticity change $\Delta\theta$□ [34,35] (see Appendix A for the detection scheme of the MOKE measurement). The F-mode oscillation is also excited by THz magnetic field along the x or y-axis. However, the magnetization deviations associated with the F-mode, $\Delta M_x$ and $\Delta M_y$, do not contribute to the MOKE in our experimental geometry, where the probe light was incident normal to the c-cut surface of $HoFeO_3$ (the xy-plane) [34,35]. In addition, the amplitude of the F-mode is much smaller than AF-mode because the F-mode resonance frequency ($\nu_F$~0.37 THz) differs from the LC resonance frequency ($\nu_{LC}$~0.56 THz).

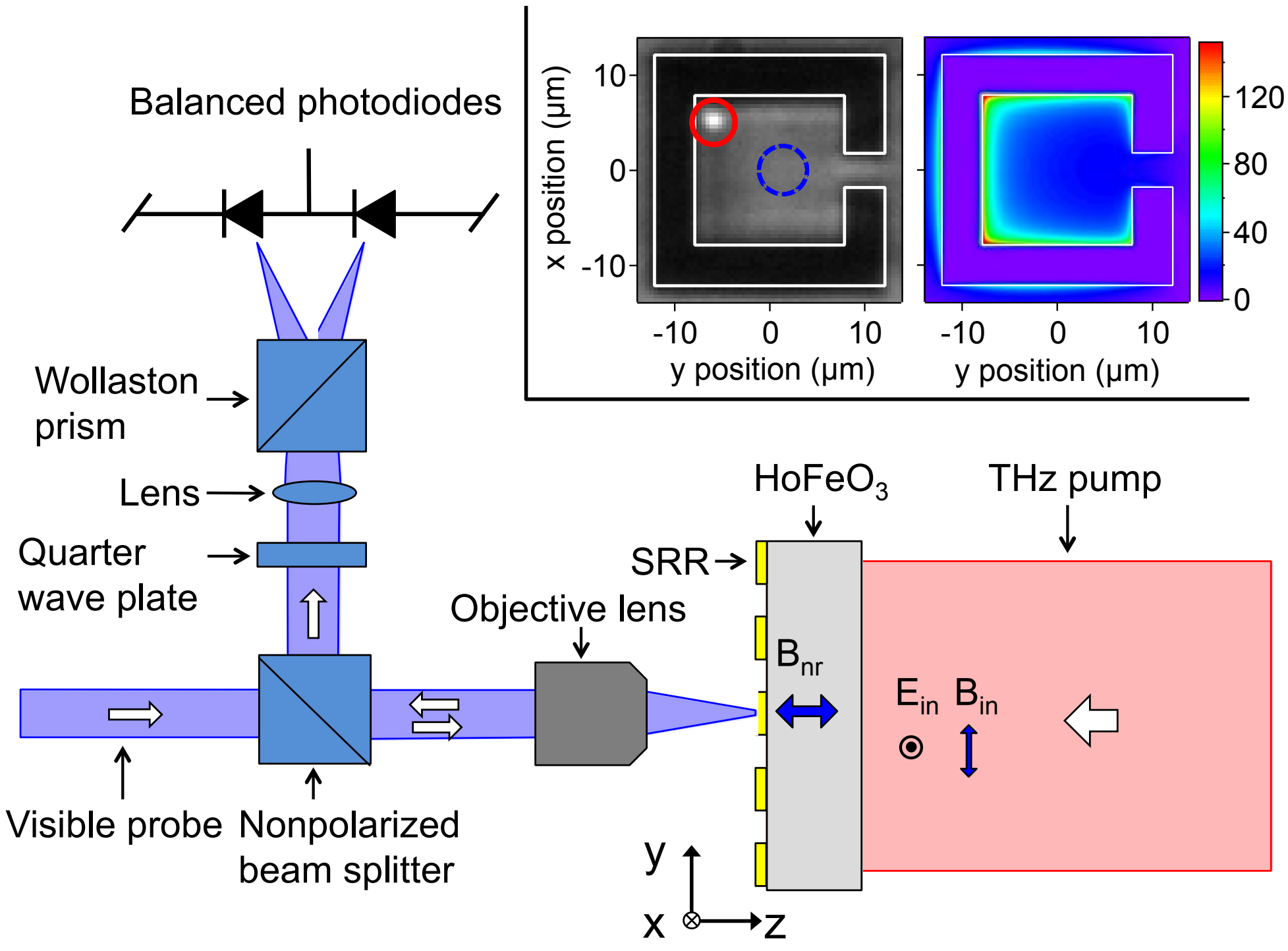

**Figure 1.** Schematic setup of THz pump-visible MOKE measurement. The left side of the inset shows the photograph of SRR fabricated on the c-cut surface of the $HoFeO_3$ crystal and the white solid line indicates the edge of the SRR. The red solid and blue dashed circles indicate the probe spots for the MOKE measurement. The right side of the inset shows the spatial distribution of the enhancement factor calculated by the FDTD method, i.e., the ratio between the Fourier amplitude at $\nu_{LC}$ of the z-component of $B_{nr}$ (at $z$=0) and the incident THz pulse $B_{in}$.

To detect the magnetization change induced only by the enhanced magnetic field, the MOKE signal just around the corner of the SRR (indicated by the red circle in figure 1's inset), where the magnetic field is enhanced 50-fold at the LC resonance frequency, was measured with a 400 nm probe pulse focused by an objective lens (spot diameter of ~1.5 μm). Furthermore, although the magnetic field reaches a maximum at the surface and decreases along the z-axis with a decay length of $l_{THz}$~5 μm, the MOKE measurement in reflection geometry, in contrast to the Faraday measurement in transmission [23], can evaluate the magnetization change induced only by the enhanced magnetic field around the sample surface since the penetration depth of 400 nm probe light for typical orthoferrites is on the order of tens of nm [35]. (The optical refractive indices of rare-earth orthoferrites in the near ultraviolet region including $HoFeO_3$ are similar to each other, regardless of the rare-earth ion species, because it is mostly determined by the strong optical absorption due to charge transfer and orbital promotion transitions inside the $FeO_6$ tetragonal cluster [35].) All experiments in this study were performed at room temperature.

## 3. Results and discussions

Figure 2(a) (upper panel) shows the calculated temporal magnetic waveform together with the incident magnetic field. The maximum peak amplitude is four times that of the incident THz pulse in the time domain and reaches 0.91 T. The magnetic field continues to ring until around 25 ps after the incident pulse has decayed away. The spectrum of the pulse shown in figure 2(c) has a peak at the LC resonance frequency ($\nu_{LC}$=0.56 THz) of the SRR, which is designed to coincide with the resonance frequency of the AF-mode ($\nu_{AF}^0$=0.575 THz). Figure 2(a) (lower panel) shows the time development of the MOKE signal $\Delta\theta$ for the highest THz excitation intensity (pump fluence $I$ of 292 μJ/cm$^2$ and maximum peak magnetic field $B_{max}$ of 0.91 T). The temporal evolution of $\Delta\theta$ is similar to that of the Faraday rotation measured in the previous study and the magnetization oscillates harmonically with a period of ~2 ps [23], implying that the THz magnetic field coherently drives the AF-mode motion.

As shown in figure 2(b), as the incident pump pulse intensity increases, the oscillation period becomes longer. The Fourier transform spectra of the MOKE signals for different pump intensities are plotted in figure 2(c). As the excitation intensity increases, the spectrum becomes asymmetrically broadened on the lower frequency side and its peak frequency becomes redshifted. Figure 2(d) plots the center-of-mass frequency (open circles) and the integral (closed circles) of the power spectrum $P(\nu)$ as a function of incident pulse fluence. The center frequency monotonically redshifts and $P(\nu)$ begins to saturate. As shown in figure 2(c), the MOKE spectra obtained at the center of the SRR (indicated in the inset of figure 1) does not show a redshift even for the highest intensity excitation, suggesting that the observed redshift originates from the nonlinearity of the precessional spin motion rather than that of the SRR response. We took the analytic signal approach (ASA) to obtain the time development of the instantaneous frequency $\nu(t)$ (figure 3(c)) and the envelope amplitude $\zeta_0(t)$ (figure 3(d)) from the measured magnetization change $\zeta(t)=\Delta M_z(t)/|\boldsymbol{M}_S|$ (figure 3(b)) (see Appendix B for the details of the analytic signal approach). As is described in the Appendix C, the MOKE signal

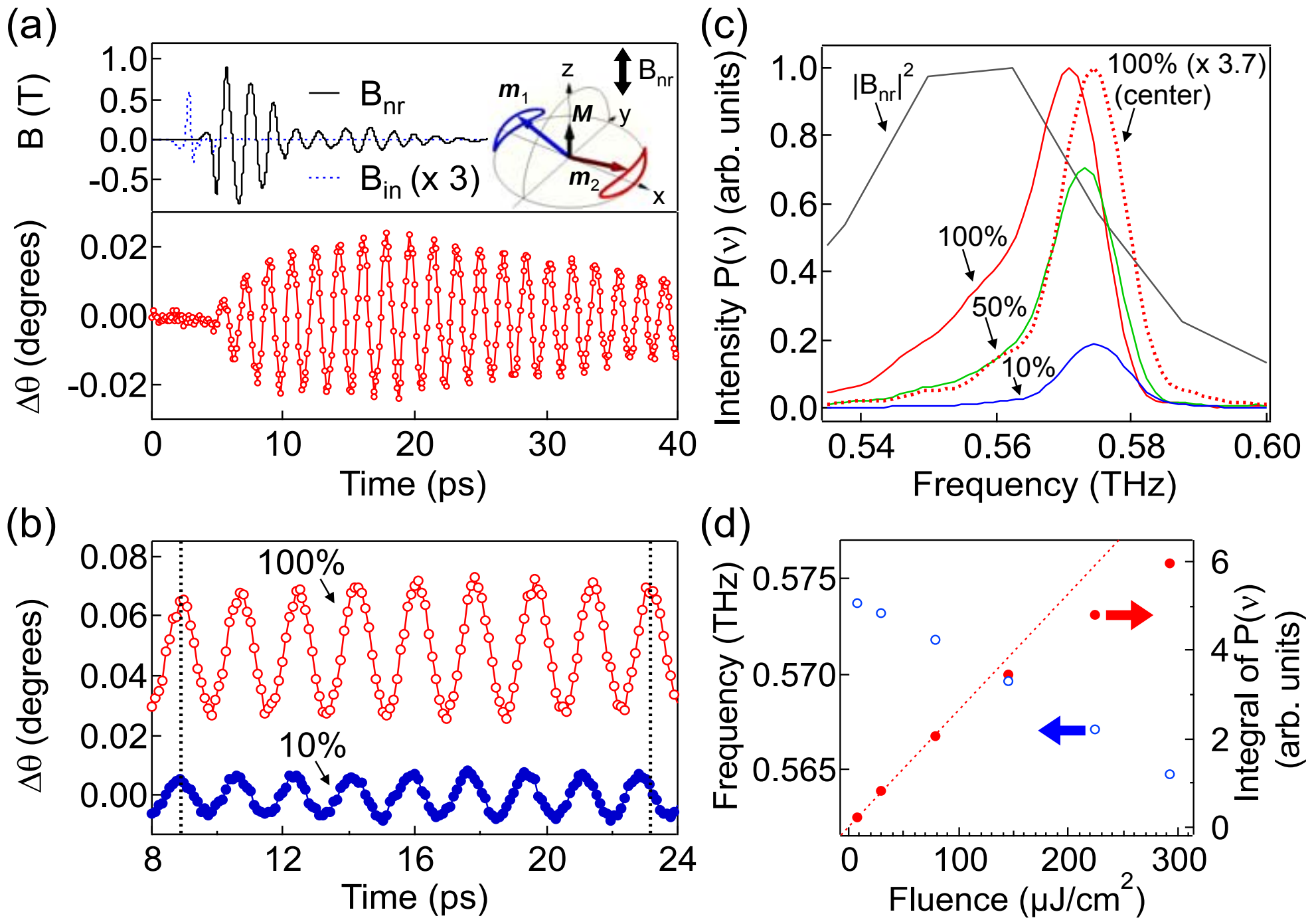

**Figure 2.** (a) Upper panel: Incident magnetic field of the THz pump pulse $B_{\mathrm{in}}$ estimated by electro-optic sampling (dashed line) and the THz magnetic near-field $B_{\mathrm{nr}}$ calculated by the FDTD method (solid line). The illustration shows the magnetization motion for the AF-mode. Lower panel: The MOKE signal for a pump fluence of 292 μJ/cm$^2$ (100%). (b) Comparison of two MOKE signals for different pump fluences, vertically offset for clarity. (c) The FFT power spectrum of the magnetic near-field $B_{\mathrm{nr}}$ (black solid line). The spectra $P(\nu)$ of the MOKE signals for a series of pump fluences obtained at the corner (solid lines) and at the center (blue dashed circle in the inset of figure 1) for a pump fluence of 100% (dashed line). Each spectrum of the MOKE signal is normalized by the peak amplitude at the corner for a pump fluence of 100%. (d) Intensity dependence of the center-of-mass frequency (open circles) and the integral (closed circles) of the $P(\nu)$.

$\Delta\theta(t)$ is calibrated to the magnetization change $\zeta(t)$ by using a linear relation, i.e., $\zeta(t)=g\Delta\theta(t)$, where $g$ (=17.8 degrees$^{-1}$) is a conversion coefficient. The time resolved experiment enables us to separate the contributions of the applied magnetic field and magnetization change to the frequency shift in the time domain. A comparison of the temporal profiles between the driving magnetic field (figure 3(a)) and the frequency evolution (figure 3(c)) shows that for the low pump fluence (10%, closed blue circles), the frequency is redshifted only when the magnetic field persists ($t$ < 25 ps), and after that, it recovers to the constant AF mode frequency ($\nu^0_{AF}$=0.575 THz). This result indicates that the signals below $t$ = 25 ps are affected by the persisting driving field and the redshift may originate from the forced oscillation. As long as the magnetic response is under the linear regime, the instantaneous frequency is independent on the pump fluence. However, for the high pump fluence (100%) a redshift (a maximum redshift of ~15 GHz relative to the constant frequency $\nu_{AF}^{0}$) appears in the delay time ($t$ < 25 ps) and even after the driving field decays away ($t$ > 25 ps) the frequency continues to be redshifted as long as the amplitude of the magnetization change is large. These results suggest that the frequency redshift in the high intensity case depends on the magnitude of the magnetization change, implying that its origin is a nonlinear precessional spin motion with a large amplitude.

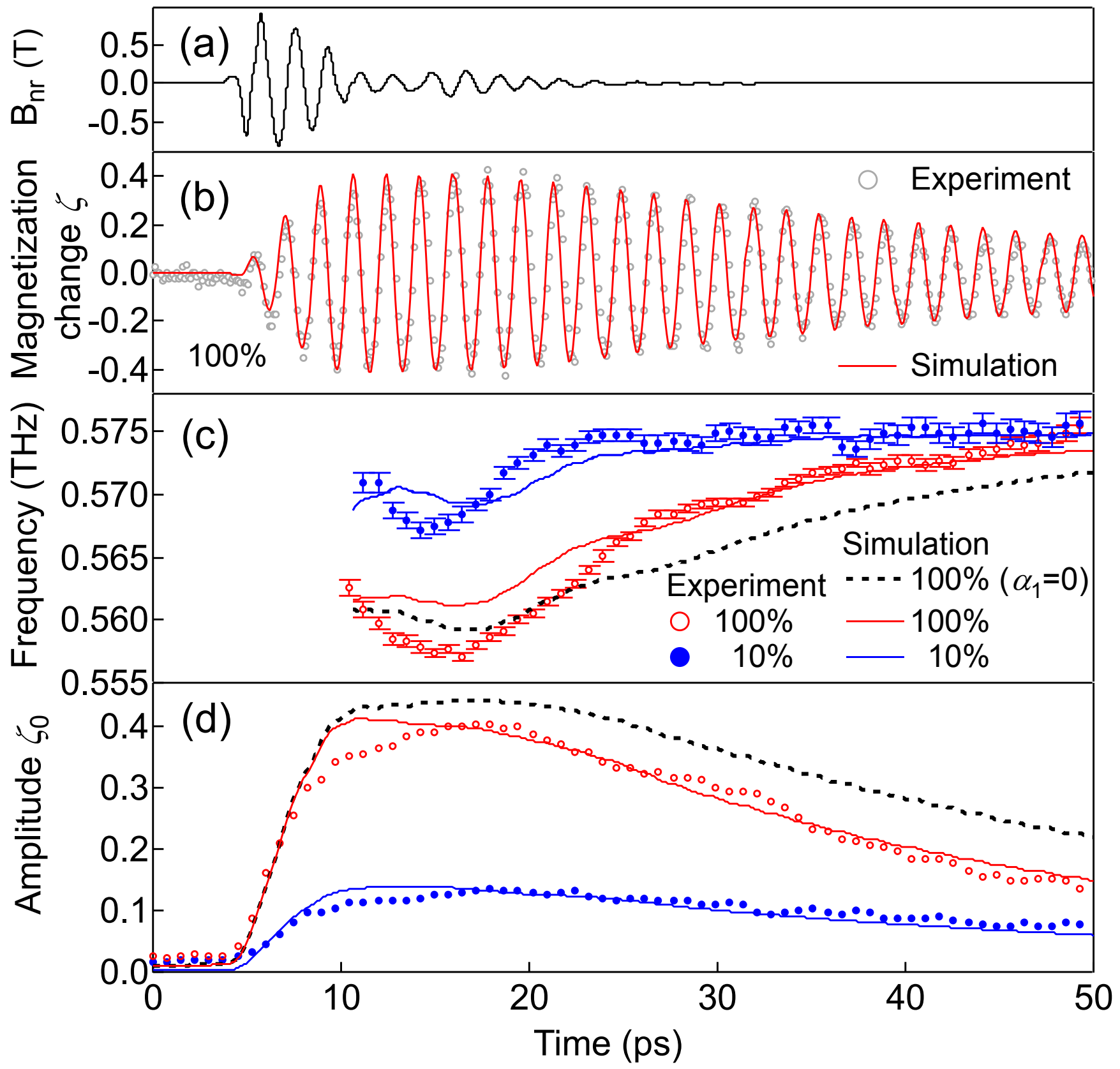

**Figure 3.** (a) FDTD calculated magnetic field $B_{nr}$ for pump fluence of 100%. (b) Temporal evolution of the magnetization change obtained from the experimental data (gray circles) and the LLG model (red line). (c) Instantaneous frequencies and (d) envelope amplitudes for pump fluences of 100% and 10% obtained by the analytic signals calculated from the experimental data (circles) and the LLG simulation with nonlinear damping parameter ($\alpha_1$=1×10$^{-3}$, solid lines) and without one ($\alpha_1$=0, dashed line).

The temperature increase due to the THz absorption (for $HoFeO_3$ $\Delta T$=1.7×10$^{-3}$ K, for gold SRR $\Delta T$=1 K) is very small (see Appendix D). In addition, the thermal relaxation of the spin system, which takes more than a nanosecond [36], is much longer than the frequency modulation decay (~50 ps) in figure 3(c). Therefore, laser heating can be ignored as the origin of the redshift.

Figure 4 shows a parametric plot of the instantaneous frequency $\nu(t)$ and envelope amplitude $\zeta_0(t)$ for the high pump fluence (100%). The instantaneous frequency shift for $t$ > 25 ps has a square dependence on the amplitude, i.e., $\nu_{AF}=\nu_{AF}^{0}(1-C\zeta_0^2)$. To quantify the relationship between the redshift and magnetization change, it would be helpful to have an analytical expression of the AF mode frequency $\nu_{AF}$ as a function of the magnetization change, which is derived from the LLG equation based on the two-lattice model [32,33]. The dynamics of the sublattice magnetizations $\boldsymbol{m}_i$ (i=1,2), as shown in the inset of figure 2(a), are described by

$$\frac{d\boldsymbol{R}_i}{dt}=-\frac{\gamma}{(1+\alpha^2)}\left(\boldsymbol{R}_i\times[\boldsymbol{B}(t)+\boldsymbol{B}_{eff,i}]-\alpha\boldsymbol{R}_i\times\left(\boldsymbol{R}_i\times[\boldsymbol{B}(t)+\boldsymbol{B}_{eff,i}]\right)\right), \quad (1)$$

where $\boldsymbol{R}_i=\boldsymbol{m}_i/m_0$ ($m_0=|\boldsymbol{m}_i|$) is the unit directional vector of the sublattice magnetizations, $\gamma$=1.76×10$^{11}$ s$^{-1}$T$^{-1}$ is the gyromagnetic ratio, $V(\boldsymbol{R}_i)$ is the free energy of the iron spin system normalized with $m_0$, and $\boldsymbol{B}_{eff,i}$ is the effective magnetic field given by $-\partial V/\partial \boldsymbol{R}_i$ (i=1,2) (see Appendix E). The second term represents the magnetization damping with the Gilbert damping constant $\alpha$□

Since $\boldsymbol{B}_{eff,i}$ depends on the sublattice magnetizations $\boldsymbol{m}_i$ and the product of these quantities appears on the right side of Eq. (1), the LLG equation is intrinsically nonlinear. If the angle of the sublattice magnetization precession is sufficiently small, Eq. (1) can be linearized and the two fixed AF- and F-modes for the weak excitation can be derived. However, as shown in figure 3(b), the deduced maximum magnetization change $\zeta$ reaches ~0.4, corresponding to precession angles of 0.25° in the xz-plane and 15° in the xy-plane. Thus, the magnetization change might be too large to use the linear approximation. For such a large magnetization motion, assuming the amplitude of the F-mode is zero and $\alpha$=0 in Eq. (1), the AF mode frequency $\nu_{AF}$ in the nonlinear regime can be deduced as

$$\nu_{AF} = \nu_{AF}^0 \frac{\sqrt{1-\zeta_0^2 \tan^2 \beta_0}}{K(D)}, \tag{2}$$

$$D = \sqrt{\frac{\zeta_0^2 (r_{AF}^2 - 1) \tan^2 \beta_0}{1-\zeta_0^2 \tan^2 \beta_0}}, \tag{3}$$

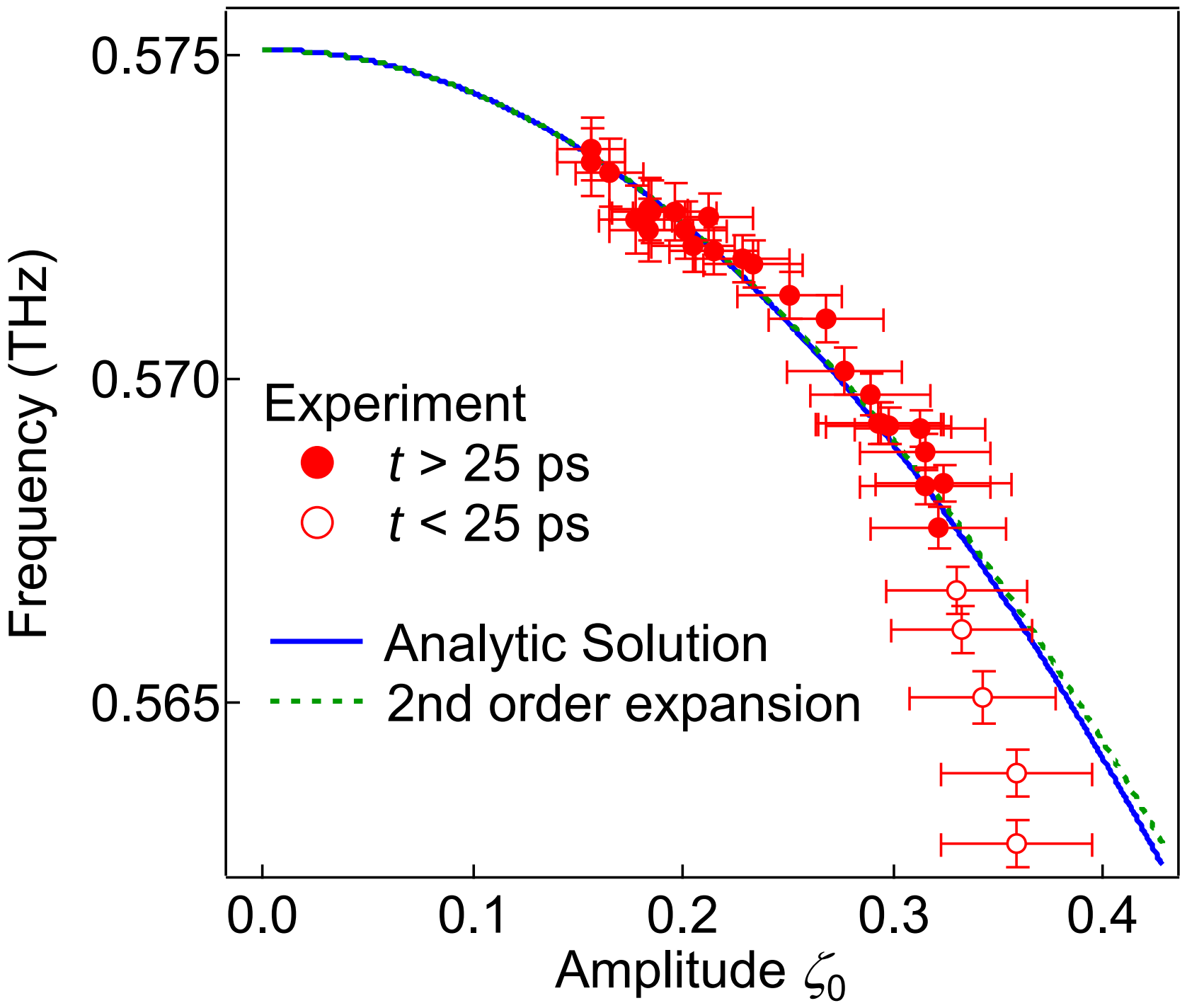

**Figure 4.** Relation between instantaneous frequency $\nu$ and envelope amplitude $\zeta_0$ obtained from the magnetization change; for $t < 25$ ps (open circles) and for $t > 25$ ps (closed circles), the analytic solution (blue line) and second order expansion of the analytic solution (green dashed line). Errors are estimated from the spatial inhomogeneity of the driving magnetic field (see Appendix H).

where $K(D)$ is the complete elliptic integral of the first kind, $r_{AF}(\approx 60)$ is the ellipticity of the sublattice magnetization precession trajectory of the AF-mode (see Appendix F), and $\zeta_0$ is the amplitude of the $\zeta(t)$. As shown in figure 4, the analytic solution can be approximated by the second order expansion $\nu_{AF} \approx \nu_{AF}^0 (1 - \tan^2 \beta_0 (r_{AF}^2 - 1) \zeta_0^2 / 4)$ and matches the observed redshift for $t > 25$ ps, showing that the frequency approximately decreases with the square of $\zeta(t)$. The discrepancy of the experimental data from the theoretical curve ($t < 25$ ps) may be due to the forced oscillation of the AF-mode caused by the driving field.

To elaborate the nonlinear damping effects, we compared the measured $\zeta(t)$ with that calculated from the LLG equation with the damping term. As shown in figures 3(c) and 3(d), the experiment for the high intensity excitation deviates from the simulation with a constant Gilbert damping (dashed lines) even in the $t > 25$ ps time region, suggesting nonlinear damping becomes significant in the large amplitude region. To describe the nonlinear damping phenomenologically, we modified the LLG equation so as to make the Gilbert damping parameter depend on the displacement of the sublattice magnetization from its equilibrium position, $\alpha(\boldsymbol{R}_i) = \alpha_0 + \alpha_1 \Delta R_i$. As shown in figures 3(b)-(d), the magnetization change $\zeta(t)$ derived with Eq. (1) (solid line) with the damping parameters ($\alpha_0 = 2.27\times10^{-4}$ and $\alpha_1 = 1\times10^{-3}$) nicely reproduces the experiments for both the high (100%) and low (10%) excitations.[1] These results suggest that the nonlinear damping plays a significant role in the large amplitude magnetization dynamics. Most plausible mechanism for the nonlinear damping is four magnon scattering process, which has been introduced to quantitatively evaluate the magnon mode instability of ferromagnet in the nonlinear response regime [37].

## 4. Conclusions

In conclusion, we studied the nonlinear magnetization dynamics of a $HoFeO_3$ crystal excited by a THz magnetic field and measured by MOKE microscopy. The intense THz field can induce the large magnetization change (~40%), and the magnetization change can be kept large enough to induce the redshift even after the field has gone, enabling us to separate the contributions of the applied magnetic field and magnetization change to the frequency shift in the time domain. The resonance frequency decreases in proportion to the square of the magnetization change. A modified LLG equation with a phenomenological nonlinear damping term quantitatively reproduced the nonlinear dynamics. This suggests that a nonlinear spin relaxation process should take place in a strongly driven regime. This study opens the way to the study of the practical limits of the speed and efficiency of magnetization reversal, which is of vital importance for magnetic recording and information processing technologies.

---

[1] The damping parameter $\alpha_0$ ($=2.27\times10^{-4}$) and conversion coefficient $g$ ($=17.8$ degrees$^{-1}$) are determined from the least-squares fit of the calculated result without the nonlinear damping parameter $\alpha_1$ to the experimental MOKE signal for the low pump fluence of 29.2 μJ/cm$^2$. The nonlinear damping parameter $\alpha_1$ ($=1\times10^{-3}$) is obtained by fitting the experimental result for the high intensity case ($I$=292 μJ/cm$^2$) with the values of $\alpha_0$ and $g$ obtained for the low excitation experiment. The estimated value of $g$ is consistent with the static MOKE measurement; the Kerr ellipticity induced by the spontaneous magnetization $\boldsymbol{M}_S$ is ~0.05 degrees ($g$~20 degrees$^{-1}$). See Appendix G for details on the static Kerr measurement.

**Acknowledgments**

We are grateful to Shintaro Takayoshi, Masahiro Sato, and Takashi Oka for their discussions with us. This study was supported by a JSPS grants (KAKENHI 26286061 and 26247052) and Industry-Academia Collaborative R&D grant from the Japan Science and Technology Agency (JST).

## Appendix A. Detection scheme of MOKE measurement

We show the details of the detection scheme of the MOKE measurement. A probe pulse for the MOKE measurement propagates along the z direction. By using the Jones vector [38], an electric field $\boldsymbol{E}_0$ of the probe pulse polarized linearly along the x-axis is described as

$$\boldsymbol{E}_0 = \begin{pmatrix} 1 \\ 0 \end{pmatrix}. \tag{A.1}$$

The probe pulse $\boldsymbol{E}_1$ reflected from the $HoFeO_3$ surface becomes elliptically polarized with a polarization rotation angle $\phi$ and a ellipticity angle $\theta$. It can be written as

$$\boldsymbol{E_1} = \mathbf{R}(-\phi)\mathbf{M}\mathbf{R}(\theta)\boldsymbol{E}_0 = \begin{pmatrix} \cos\theta\cos\phi - i\sin\theta\sin\phi \\ \cos\theta\sin\phi + i\sin\theta\cos\phi \end{pmatrix}, \tag{A.2}$$

where $\mathbf{M}$ is the Jones matrix describing $\pi$□□ phase retardation of the y component with respect to the x component

$$\mathbf{M} = \begin{pmatrix} 1 & 0 \\ 0 & -i \end{pmatrix}, \tag{A.3}$$

and $\mathbf{R}(\psi)$ is the rotation matrix

$$\mathbf{R}(\psi) = \begin{pmatrix} \cos\psi & \sin\psi \\ -\sin\psi & \cos\psi \end{pmatrix}. \tag{A.4}$$

The reflected light passes through the quarter wave plate, which is arranged such that its fast axis is tilted by an angle of 45° to the x-axis. The Jones matrix of the wave plate is given by

$$\mathbf{R}\left(-\frac{\pi}{4}\right)\mathbf{M}\mathbf{R}\left(\frac{\pi}{4}\right). \tag{A.5}$$

Thus, the probe light $\boldsymbol{E}_2$ after the quarter wave plate is described as follows,

$$\boldsymbol{E_2} = \begin{pmatrix} E_{2,x} \\ E_{2,y} \end{pmatrix} = \mathbf{R}\left(-\frac{\pi}{4}\right)\mathbf{M}\mathbf{R}\left(\frac{\pi}{4}\right)\boldsymbol{E_1}$$

$$= \frac{1}{2}\begin{pmatrix} \cos(\theta+\phi) - \sin(\theta-\phi) + i(-\cos(\theta-\phi) + \sin(\theta+\phi)) \\ \cos(\theta-\phi) + \sin(\theta+\phi) + i(\cos(\theta+\phi) + \sin(\theta-\phi)) \end{pmatrix}. \quad (A.6)$$

The Wollaston prism after the quarter wave plate splits the x and y-polarization components of the probe light $\boldsymbol{E}_2$. The spatially separated two pulses are incident to the balanced detector and the detected probe pulse intensity ratio of the differential signal to the total corresponds to the Kerr ellipticity angle as follows,

$$\frac{\langle |E_{2,x}|^2\rangle - \langle |E_{2,y}|^2\rangle}{\langle |E_{2,x}|^2\rangle + \langle |E_{2,y}|^2\rangle} = -\sin 2\theta. \quad (A.7)$$

In the main text, we show the Kerr ellipticity change $\Delta\theta = \theta_{w} - \theta_{wo}$, where the ellipticity angles ($\theta_{w}$ and $\theta_{wo}$) are respectively obtained with and without the THz pump excitation.

**Appendix B. Analytic signal approach and short time Fourier transform**

The Analytic signal approach (ASA) allows the extraction of the time evolution of the frequency and amplitude by a simple procedure and assumes that the signal contains a single oscillator component. In our study, we measure only the MOKE signal originating from the AF-mode and it can be expected that the single oscillator assumption is valid. In the ASA, the time profile of the magnetization change $\zeta(t)$ is converted into an analytic signal $\psi(t)$, which is a complex function defined by using the Hilbert transform [39];

$$\psi(t) = \zeta_0(t)\exp(i\phi(t)) = \zeta(t) + i\,\tilde{\zeta}(t), \quad (B.1)$$

$$\tilde{\zeta}(t) = \frac{1}{\pi}\, p\int_{-\infty}^{\infty} \frac{\zeta(t)}{t-\tau}\, \mathrm{d}\tau. \quad (B.2)$$

where the $p$ is the Cauthy principal value. The real part of $\psi(t)$ corresponds to $\zeta(t)$. The real function $\zeta_0(t)$ and $\phi(t)$ represent the envelope amplitude and instantaneous phase of the magnetization change. The instantaneous frequency $\omega(t)(=2\pi\nu(t))$ is given by $\omega(t) = \mathrm{d}\phi(t)/\mathrm{d}t$. In the analysis, we averaged $\zeta_0(t)$ and $\omega(t)$ over a ten picosecond time range.

To confirm whether the ASA gives appropriate results, as shown in figure B.1 we compare them with those obtained by the short time Fourier transform (STFT). As shown in figure B.1(a), the time-frequency plot shows only one oscillatory component of the AF-mode. As shown in figures B.1(b) and (c), the instantaneous frequencies and amplitudes obtained by the ASA and the STFT are very similar. Because the ASA provides us the instantaneous amplitude with a simple procedure, we showed the time evolutions of frequency and amplitude derived by the ASA in the main text.

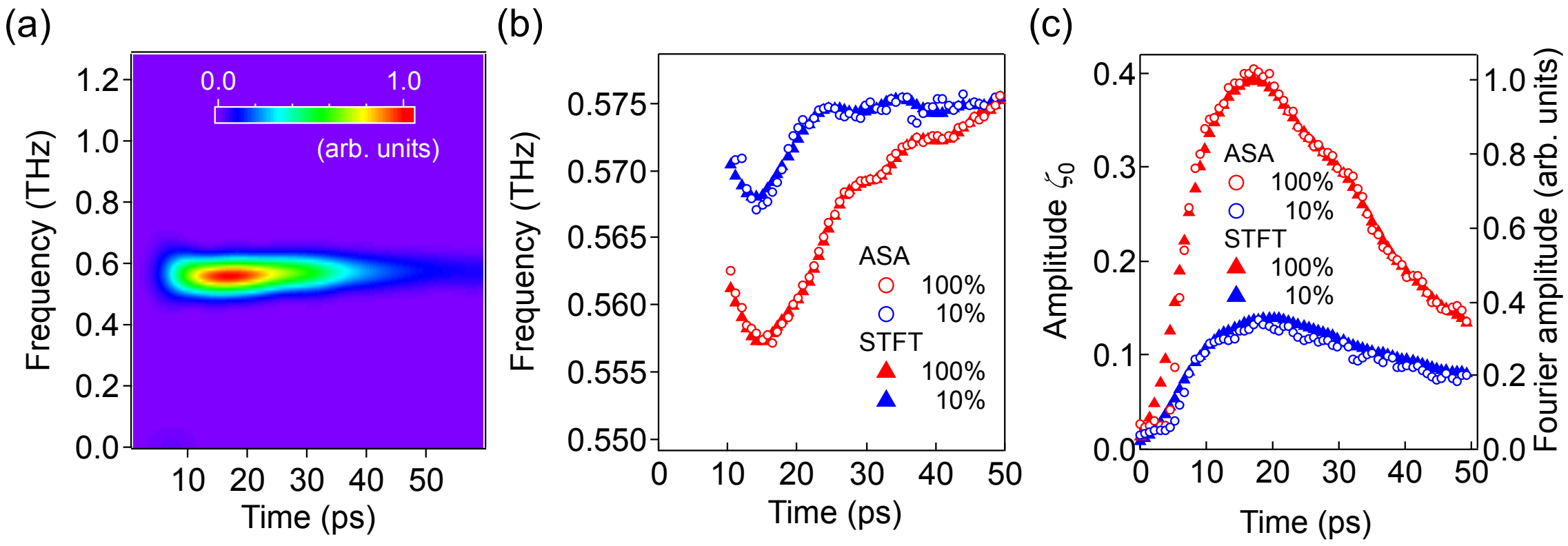

**Figure B.1.** (a) Time-dependence of the power spectrum of the magnetization oscillation for the highest THz excitation ($I$=292 μJ/cm$^2$) obtained by the STFT. Comparison of (b) instantaneous frequencies and (c) amplitudes obtained by the ASA and STFT with a time window with FWHM of 10 ps.

### Appendix C. Determination of conversion coefficient *g* and linear damping parameter $\alpha_0$

The conversion coefficient $g$ and the linear damping parameter $\alpha_0(=\alpha)$ in Eq. (1) are determined by fitting the experimental MOKE signal $\Delta\theta(t)$ for the low pump fluence of 29.2 μJ/cm$^2$ with the LLG calculation of the magnetization change $\zeta(t)$. Figure C.1 shows the MOKE signal $\Delta\theta(t)$ (circle) and the calculated magnetization change $\zeta(t)$ (solid line). From the least-squares fit of the calculated result to the experiment by using a linear relation, i.e., $\zeta(t)=g\Delta\theta(t)$, we obtained the parameters $g$(=17.8 degrees$^{-1}$) and $\alpha_0(=2.27\times10^{-4})$.

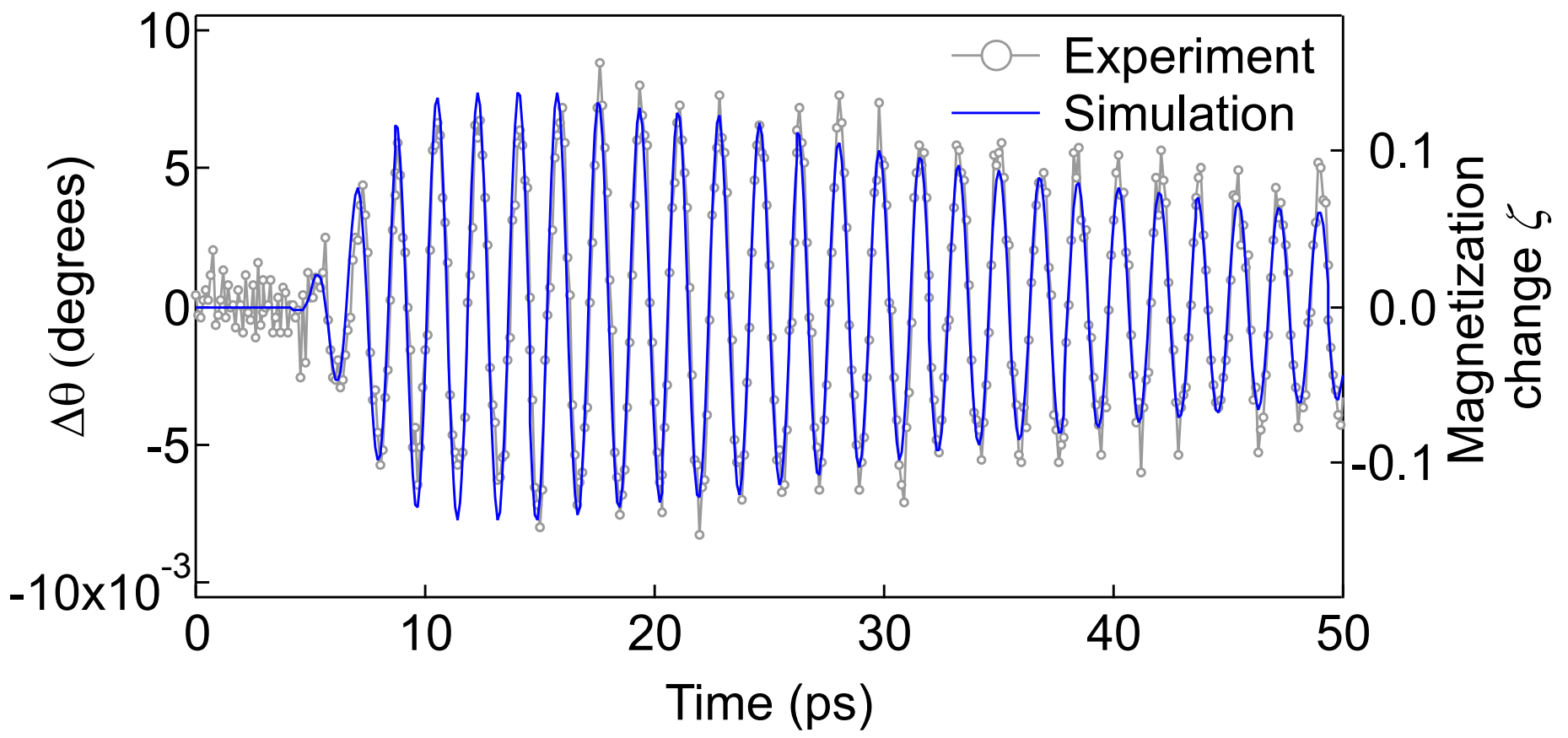

**Figure C.1.** Experimentally observed MOKE signal $\Delta\theta$ (circle) and LLG simulation result of the magnetization change $\zeta$ (solid line) for the pump fluence of 29.2 μJ/cm$^2$.

**Appendix D. Laser heating effect**

The details of the calculation of the temperature change are as follows:

For $HoFeO_3$:

The absorption coefficient $\alpha_{abs}$ of $HoFeO_3$ at 0.5 THz is ~4.4 cm$^{-1}$ [40]; the fluence $I_{HFO}$ absorbed by $HoFeO_3$ can be calculated as $I_{HFO}=I(1-\exp(-\alpha_{abs}d))$, where $d$ (=145 μm) is the sample thickness and $I$ is the THz pump fluence. For the highest pump fluence, $I$=292 μJ/cm$^2$, $I_{HFO}$ is 18.1 μJ/cm$^2$. Since the sample thickness is much smaller than the penetration depth, $d \ll \alpha_{abs}^{-1}$, we assume that the heating of the sample due to the THz absorption is homogeneous. By using the heat capacity $C_p$ of 100 J mol$^{-1}$ K$^{-1}$ [27], and the molar volume $v$ of ~1.4×10$^2$ cm$^3$/mol [27], the temperature change $\Delta T$ can be estimated as $\Delta T=I_{HFO}v/C_pd$ ~1.7×10$^{-3}$ K.

For gold resonator (SRR):

The split ring resonator has an absorption band (center frequency ~0.56 THz, band width ~50 GHz) originated from the LC resonance (figure 2(c)). Assuming the SRR absorbs all incident THz light in this frequency band, the absorbed energy accounts for 3 % of the total pulse energy. Hence, for the highest THz pump fluence, $I$=292 μJ/cm$^2$, the fluence absorbed by the SRR is $I_{gold}$=8.76 μJ/cm$^2$. By using the heat capacity $C_p$ of 0.13 J g$^{-1}$ K$^{-1}$ [41], the number of the SRRs per unit area $N$ of 4×10$^4$ cm$^{-2}$, and the mass of the SRR $m$ of 1.6×10$^{-9}$ g, the temperature change

$\Delta T$ can be estimated as□$\Delta T = I_{gold}/C_p N m \sim 1$ K

**Appendix E. Free energy of $HoFeO_3$**

The free energy $F$ of the iron spin ($Fe^{3+}$) system based on the two-lattice model is a function of two different iron sublattice magnetizations $\boldsymbol{m}_i$, and composed of the exchange energy and one-site anisotropy energy [32,33]. The free energy normalized by the sublattice magnetization magnitude, $V=F/m_0$ ($m_0=|\boldsymbol{m}_i|$), can be expanded as a power series in the unit directional vector of the sublattice magnetizations, $\boldsymbol{R}_i=\boldsymbol{m}_i/m_0=(X_i,Y_i,Z_i)$. In the magnetic phase $\Gamma_4$ ($T$ > 58K), the normalized free energy is given as follows [32,33]:

$$V=E\boldsymbol{R}_1\cdot\boldsymbol{R}_2+D(X_1Z_2-X_2Z_1)-A_{xx}(X_1^2+X_2^2)-A_{zz}(Z_1^2+Z_2^2), \quad \text{(E.1)}$$

where $E$(=6.4×10$^2$ T) and $D$(=1.5×10 T) for $HoFeO_3$ are respectively the symmetric and antisymmetric exchange field [42]. $A_{xx}$ and $A_{zz}$ are the anisotropy constants. As mentioned in Appendix F, the temperature dependent values of the anisotropy constants can be determined from the antiferromagnetic resonance frequencies. The canting angle of $\boldsymbol{R}_i$ to the x-axis $\beta_0$ under no magnetic field is given by

$$\tan 2\beta_0=\frac{D}{E+A_{xx}-A_{zz}}. \quad \text{(E.2)}$$

**Appendix F. Linearized resonance modes and anisotropy constants ($A_{xx}$ and $A_{zz}$)**

The nonlinear LLG equation of Eq. (1) can be linearized and the two derived eigenmodes correspond to the AF and F-mode. The sublattice magnetization motion for each mode is given by the harmonic oscillation of mode coordinates; for the AF-mode ($Q_{AF}$, $P_{AF}$)=(($X_1-X_2$) $\sin\beta_0$ +($Z_1+Z_2$) $\cos\beta_0$ , $Y_1-Y_2$), and for the F-mode ($Q_F$, $P_F$)=(($X_1+X_2$)$\sin\beta_0$−($Z_1-Z_2$)$\cos\beta_0$, $Y_1+Y_2$),

$$Q_{AF}=A_{AF}\cos\omega_{AF}t, \quad \text{(F.1)}$$

$$P_{AF}=A_{AF}r_{AF}\sin\omega_{AF}t, \quad \text{(F.2)}$$

$$Q_F=A_F\cos\omega_F t, \quad \text{(F.3)}$$

$$P_F=A_F r_{AF}\sin\omega_F t, \quad \text{(F.4)}$$

where $A_{\mathrm{AF,F}}$ represents the amplitude of each mode. $\omega_{\mathrm{AF,F}}$, and $r_{\mathrm{AF,F}}$ are the resonance frequencies and ellipticities, which are given by

$$\omega_{\mathrm{AF}}=\gamma\sqrt{(b+a)(d\text{-}c)}, \tag{F.5}$$

$$\omega_{\mathrm{F}}=\gamma\sqrt{(b\text{-}a)(d+c)}, \tag{F.6}$$

$$r_{\mathrm{AF}}=\gamma\sqrt{\frac{(d-c)}{(b+a)}}, \tag{F.7}$$

$$r_{\mathrm{F}}=\gamma\sqrt{\frac{(d+c)}{(b\text{-}a)}}, \tag{F.8}$$

where $\gamma$=1.76×10$^{11}$ s$^{-1}$T$^{-1}$ is the gyromagnetic ratio, and

$$a=-2A_{\mathrm{xx}}\cos^2\beta_0-2A_{\mathrm{zz}}\sin^2\beta_0-E\cos 2\beta_0-D\sin 2\beta_0, \tag{F.9}$$

$$b=E, \tag{F.10}$$

$$c=2A_{\mathrm{xx}}\cos 2\beta_0-2A_{\mathrm{zz}}\cos 2\beta_0+E\cos 2\beta_0+D\sin 2\beta_0, \tag{F.11}$$

$$d=-E\cos 2\beta_0-D\sin 2\beta_0. \tag{F.12}$$

Substituting the literature values of the exchange fields ($E$=6.4×10$^2$ T and $D$=1.5×10 T [42]) and the resonance frequencies at room temperature ($\omega_{\mathrm{AF}}/2\pi$=0.575 THz and $\omega_{\mathrm{F}}/2\pi$=0.37 THz) to Eqs. (F.5) and (F.6), $A_{\mathrm{xx}}$ and $A_{\mathrm{zz}}$ can be determined to 8.8×10$^{-2}$ T and 1.9×10$^{-2}$ T.

**Appendix G. MOKE measurement for the spontaneous magnetization**

Figure G.1 shows time-development of the MOKE signals for the different initial condition with oppositely directed magnetization. We applied the static magnetic field (~0.3 T) to saturate the magnetization along the z-axis before the THz excitation. The spontaneous magnetization of single crystal $HoFeO_a$ can be reversed by the much smaller magnetic field (~0.01 T) because of the domain wall motion [27]. Then, we separately measured the static Kerr ellipticity angle $\theta$□□□□□□ and THz induced ellipticity change $\Delta\theta$ for different initial magnetization $M_z$=±Ms without the static magnetic field□ In figure G.1 we plot the summation of the time resolved MOKE signal $\Delta\theta$□and the static Kerr ellipticity $\theta$□□□□□□ The sings of the ellipticity offset angle

$\theta$□□□□□□ for the different spontaneous magnetization (±$M_S$) are different and their magnitudes are ~0.05 degrees. The conversion coefficient $g$(=1/$\theta$~□/0.05 degrees) is estimated to be ~20 degrees$^{-1}$, which is similar to the value determined by the LLG fitting (~17.8 degrees$^{-1}$). In the case of the AF-mode excitation, the phases of the magnetization oscillations are in-phase regardless of the direction of the spontaneous magnetization $\boldsymbol{M}$=±Ms, whereas they are out-of-phase in the case of the F-mode excitation. We can explain this claim as follows: In the case of AF-mode excitation, the external THz magnetic field is directed along the z-direction as shown in the inset of figure 2(a), the signs of the torques acting on the sublattice magnetization $\boldsymbol{m}_i$ ($i$=1,2) depends on the direction of $\boldsymbol{m}_i$, however, the resultant oscillation of the macroscopic magnetization $\boldsymbol{M}= \boldsymbol{m}_1+\boldsymbol{m}_2$ along the z-direction has same phase for the different initial condition $\boldsymbol{M}$=±Ms. In the case of the F-mode excitation with the external THz magnetic field along the x or y-direction, the direction of the torques acting on the magnetization $\boldsymbol{M}$ depends on the initial direction and the phase of the F-mode oscillation changes depending on the sign of the spontaneous magnetization ±Ms.

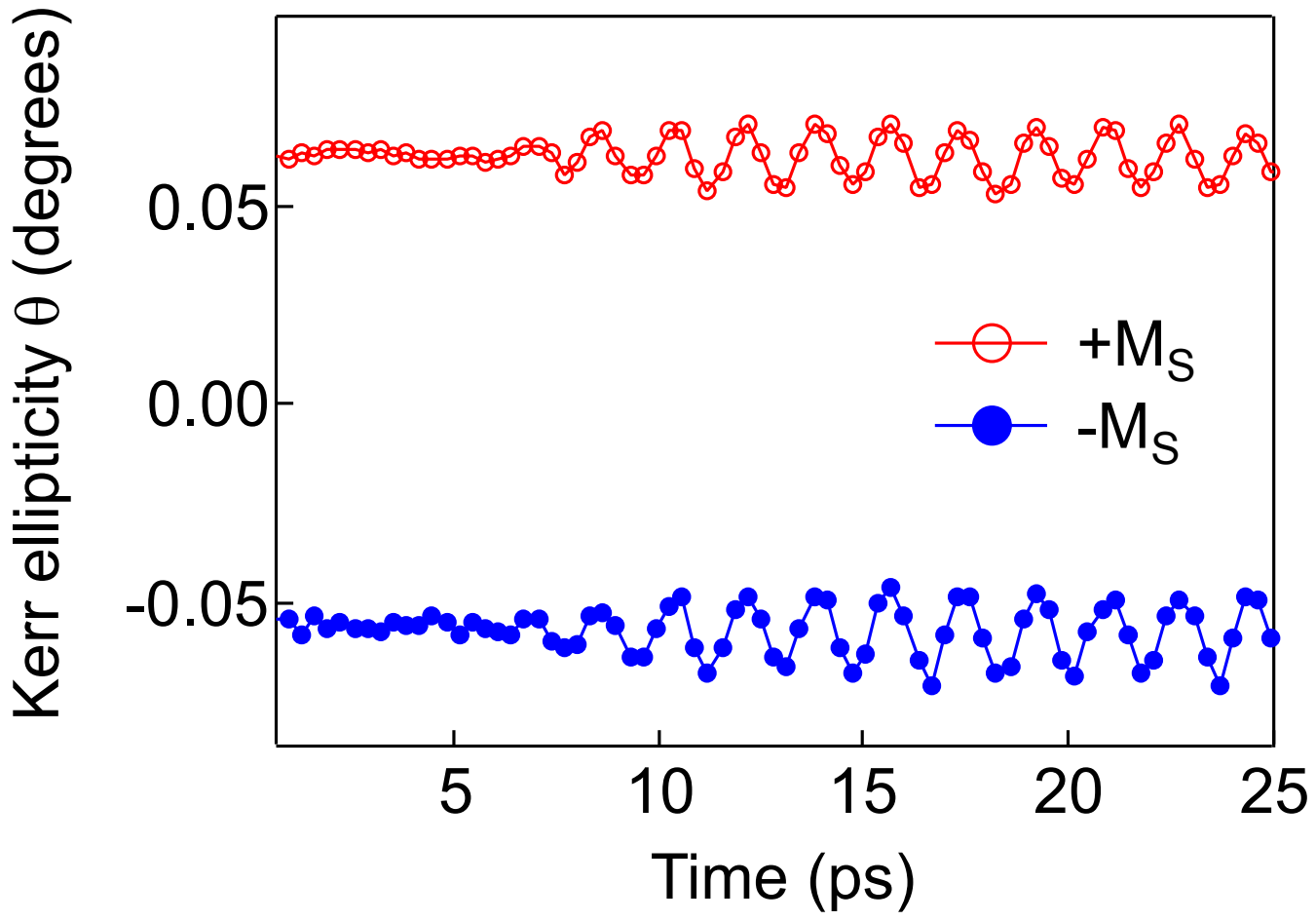

**Figure G.1.** The MOKE signals, the temporal change of the Kerr ellipticity $\theta$, measured for different initial conditions with oppositely directed magnetizations.

## Appendix H. Influence of the spatial distribution of magnetic field on magnetization change

As shown in the inset of figure 1, the pump magnetic field strongly localizes near the metallic arm of the SRR and the magnetic field strength significantly depends on the spatial position $\boldsymbol{r}$ within the probe pulse spot area. The intensity distribution of the probe pulse $I_{probe}(\boldsymbol{r})$ has an elongated Gaussian distribution with spatial widths of 1.1 μm along the x-axis and 1.4 μm along the y-axis [full width at half maximum (FWHM) intensity]. The maximum magnetic field is 1.2 times larger than the minimum one in the spot diameter, causing the different magnetization change dynamics at different positions. To take into account this spatial inhomogeneity to the simulation, the spatially weighted average of magnetization change $\bar{\zeta}(t)$ has to be calculated as follows:

$$\bar{\zeta}(t)=\frac{\int \zeta(\boldsymbol{r},t)I_{\mathrm{probe}}(\boldsymbol{r})\mathrm{d}\boldsymbol{r}}{\int I_{\mathrm{probe}}(\boldsymbol{r})d\boldsymbol{r}}\ , \tag{H.1}$$

where $\zeta(\boldsymbol{r},t)$ is a magnetization change at a position $\boldsymbol{r}$ and time $t$.

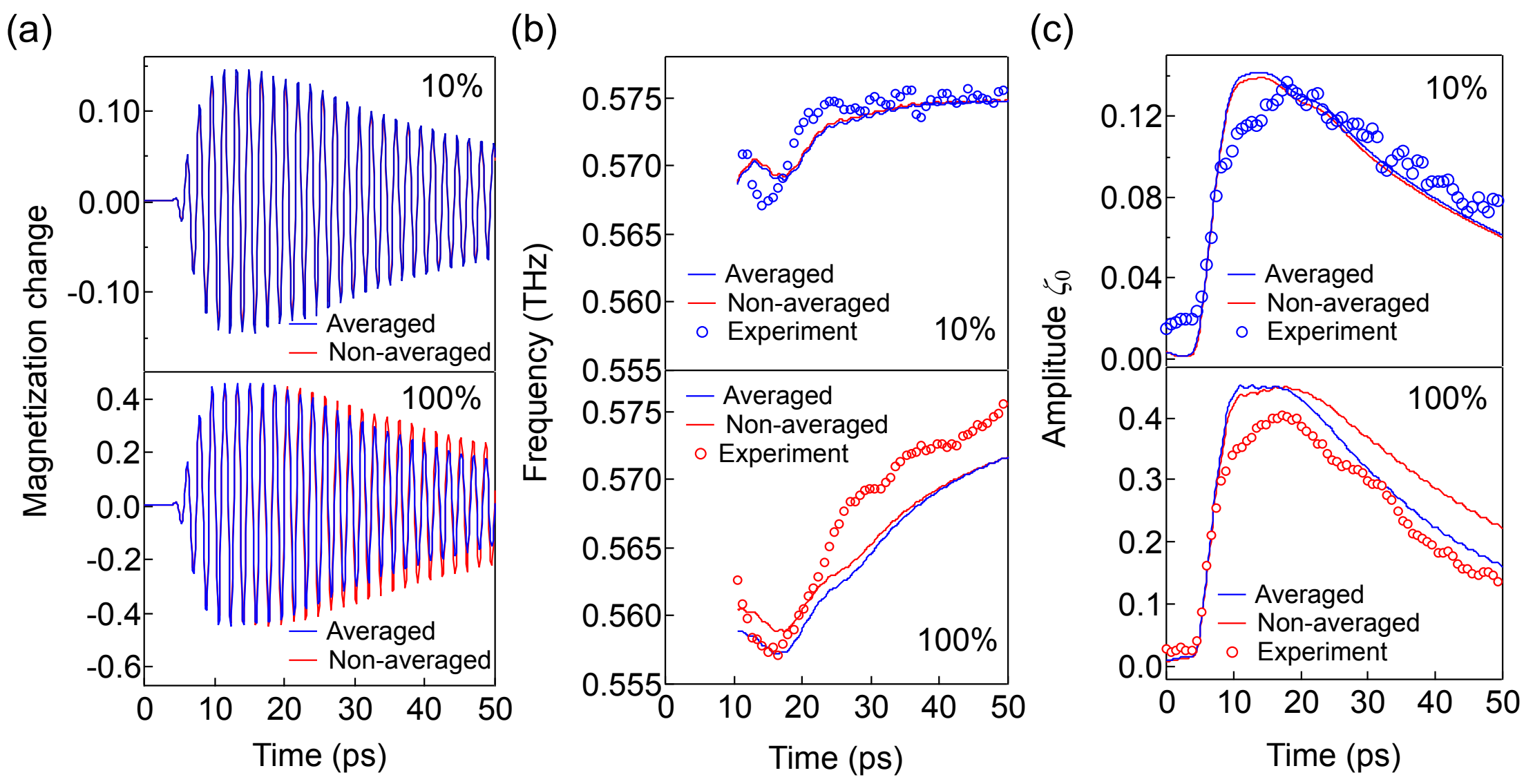

**Figure H.1.** Comparison of the spatially averaged and non-averaged magnetization change for the different pump fluences of 10% and 100%. (a) Temporal evolutions of the magnetization change, (b) instantaneous frequencies and (c) normalized envelope amplitudes. Open circles show the experimental results.

Figure H.1(a) shows the simulation result of the spatially averaged magnetization change $\bar{\zeta}(t)$ and the non-averaged $\zeta(\boldsymbol{r}_0,t)$ without the nonlinear damping term ($\alpha_1$=0), where $\boldsymbol{r}_0$ denotes the peak position of $I_{\mathrm{probe}}(\boldsymbol{r})$. For the low excitation intensity (10%), $\bar{\zeta}(t)$ is almost the same as $\zeta(\boldsymbol{r}_0,t)$ as shown in figure H.1(a). On the other hand, for the high excitation intensity, the spatial inhomogeneity of magnetization change dynamics induces a discrepancy between the $\bar{\zeta}(t)$ and $\zeta(\boldsymbol{r}_0,t)$. This discrepancy is caused by the quasi-interference effect between the magnetization dynamics with different frequencies and amplitudes at different positions. Figures H.1(b) and (c) show the instantaneous frequency and envelope amplitude obtained from the data shown in figure H.1(a) by using analytic signal approach with the experimental result. For the averaged magnetization change, the frequency redshift is more emphasized (figure H.1(b)) and the decay time becomes shorter (figure H.1(c)). Nonetheless, neither spatially averaged nor non-averaged simulation reproduces the experimental result of the instantaneous frequency (figure H.1(b)) without nonlinear damping term.